\documentclass[prl, twocolumn, showpacs]{revtex4}
\usepackage{amssymb}
\usepackage{amsmath}
\usepackage{epsfig}
\usepackage{bm}
\usepackage{color}

\begin{document}

\title{ The Hall effect in ballistic flow of two-dimensional interacting particles }
\author{P. S. Alekseev and M. A. Semina }
\affiliation{ Ioffe  Institute,  194021  St.~Petersburg, Russia }

\begin{abstract}
In high-quality solid-state systems at low temperatures, the
hydrodynamic or the ballistic regimes of heat and charge transport
are realized in the electron and  the phonon systems. In these
regimes, the thermal and the electric conductance of the sample can
reach abnormally large magnitudes. In this paper, we study the Hall effect in a
system of interacting two-dimensional charged  particles in a ballistic
regime. We demonstrated that the Hall electric field  is caused by a
change in the densities of particles due to the effect of external
fields on their free motions between the sample edges.
In one-component (electron or hole) systems the Hall coefficient turns out to one half compared with
 the one in conventional disordered Ohmic samples.
 This result is consistent with the recent experiment
 on measuring of the Hall resistance in ultra-high-mobility GaAs quantum wells.
In two-component electron-hole systems  the Hall
electric field depends linearly on the difference between the
concentrations of electrons and holes near the charge neutrality point (the
equilibrium electron and hole  densities coincide)
 and  saturates to the Hall field of a one-component system
 far from the charge
neutrality point. We also studied  the
corrections to magnetoresistance and the Hall
  electric field
  due to inter-particle scattering being a precursor of forming a viscous flow.
  For the samples shorter than the inter-particle scattering length, the obtained corrections govern
  the dependencies of  magnetoresistance and the Hall
  field on temperature.


\end{abstract}

\maketitle

\section{ Introduction }
In   novel high-quality  nanostructures and bulk materials
  extremely small densities  of defects can be achieved.
At low temperatures the electron mean free paths relative to scattering on
 disorder and on phonons in such material become very long.
  In this connection, the hydrodynamic and the ballistic regimes  of
transport can be realized in mesoscopic or even macroscopic samples.
In 1960-1970s the theory of the hydrodynamic regime of electron heat
and charge transport was developed for bulk metals  by R.~N.~Gurzhi and coauthors \cite{Gurzhi_rev}.
The ballistic electron transport of 2D electrons in semiconductor
quantum wells  was extensively studied theoretically and experimentally in 1980-1990s in several groups \cite{rev_bal}.
 In recent decade the bright evidences of realization of the hydrodynamic
 and the ballistic regimes of transport were discovered
 in several novel materials: high-mobility GaAs quantum wells,
 single-layer  graphene, 3D Weyl semimetals
\cite{exp_hydgr_1,exp_hydgr_1_2,exp_hydgr_1_3,exp_hydgr_1_4,exp_hydgr_1_ov,
exp_hydgr_2,exp_hydgr_3,exp_hydgr_4,exp_hydgr_4_2,gr_neg_MR,
Gusev_1,Gusev_2,Gusev_3,
exp_gr_1,exp_gr_2,exp_gr_3,exp_gr_4}.
 Theory of hydrodynamic and ballistic transport
 in solids has been developed in the last years in many different directions:
 both aimed for  explaining recent experiments as well as
 in areas not directly related to recent experiments
 \cite{hydr_tr_th_2,hydr_tr_th_3,hydr_tr_th_4,
hydr_tr_th_5,hydr_tr_th_6,hydr_tr_th_7,hydr_tr_th_8,
hydr_tr_th_9,hydr_tr_th_10,hydr_tr_th_10_2,hydr_tr_th_11,
hydr_tr_th_11_2,hydr_tr_th_11_3,hydr_tr_th_12,we_hd,
we_hd_2,we_hd_3,we_hd_4,we_hd_5,we_hd_5_2,th_therm_gr_hydr_1,
th_therm_gr_hydr_2,
paper_mail,paper_rec_1,Hall_visc,ours_bal,Kiselev,
recentest,recentest2,recentest3,new,new_2}.

The giant negative magnetoresistance effect is considered to be one
of the main evidences of realization
 of the hydrodynamic regime of charge transport.
 It was observed in  high-mobility GaAs quantum wells, in the  3D Weyl semimetal WP$_2$,
 and, very recently, in single-layer graphene
 \cite{exp_hydgr_1,exp_hydgr_1_2,exp_hydgr_1_3,exp_hydgr_1_4,
 exp_hydgr_1_ov,exp_hydgr_2,exp_hydgr_3,gr_neg_MR,Gusev_1}.
 The giant negative magnetoresistance often consists of a
  temperature-dependent wide peak with a large amplitude
  and of a temperature-independent  small narrow peak.
  The temperature-dependent part of the giant negative magnetoresistance
   was explained as the result of forming the viscous electron fluid
  and the magnetic field dependence of the electron viscosity \cite{hydr_tr_th_9}.
An explanation of the temperature-independent part
was proposed in   Ref.~\cite{ours_bal} within the model
 of ballistic transport of 2D interacting electrons.
 In was noted in   Ref.~\cite{ours_bal} that a small external magnetic field
  leads to the  increase of the average free path
  of ballistic electrons in a long sample.
  This fact results in a small negative magnetoresistance,
 which can be temperature-independent for not very long samples,
 where  the maximum length of ballistic trajectories is restricted
 by the sample size.

 For identification of the ballistic and the hydrodynamic
 regimes of electron transport, are important not only measurements
  of magnetoresistance and the size dependencies of
  resistance  in zero magnetic field.
  Studies of the Hall effect
   are also of great importance \cite{gr_neg_MR,Gusev_3,Hall_visc,new,new_2}.
 In Refs.~\cite{gr_neg_MR,Gusev_3} the Hall resistance
 in best-quality graphene and ultra-high mobility GaAs
 quantum wells was measured at the conditions when the hydrodynamic
 and the ballistic regime were apparently realized.
 Substantial deviations of the Hall resistance from its
  usual  value for a long Ohmic disordered sample were observed.
 In Ref.~\cite{Hall_visc} the crossover between
  the hydrodynamic and the ballistic regimes of transport,
  in particular, evolution of the longitudinal and the Hall resistances,
 were studied for 2D electrons in a long sample
  by numerical solution of the kinetic equation.
However, in that work the specific mechanisms of the Hall effect
  in the ballistic regime and the role of
  the electron-electron scattering in this regime
  were not clarified.

  In Refs.~\cite{new,new_2} the Hall effect was theoretically studied
  for a system of   a 2D interacting electrons in a weak disorder.
  The main attention was paid to the regime
  of moderate magnetic field corresponding to the cyclotron radius
  of the order of the sample width.
  It was demonstrated that the curvature
  of the Hall electric field  in the center
  of the sample can be used to distinguish
 the Ohmic, ballistic and hydrodynamic regimes \cite{new}.
    In Ref.~\cite{new_2} a method of experimental measurements
    of a generalized Hall viscosity were proposed
    at  the crossover between the ballistic and
     the hydrodynamic regime of transport in the Hall and Corbino samples.

In this paper we develop a theory of the Hall effect in
  one-component and two-component conduction systems
  of interacting particles in ballistic samples
   at small magnetic fields \cite{rep}.
We study the Hall effect
  for low magnetic field when the magnetic field term
 in the kinetic equation can be treated as a perturbation.
     Due to
  the kinematic effect of  the external fields
    on the ballistic  trajectories, the
  electron and hole densities become inhomogeneous
   and not equal one to another
  that leads to arising the Hall electric field.
The resulting  Hall coefficient turns out to be one half
 of the Hall coefficient $R_H^0 $ of the conventional
 Ohmic bulk conductor at zero temperature.
We demonstrate that the obtained result
 is consistent with the experimental data \cite{Gusev_3}.

  For two-component electron-hole systems, we studied the Hall effect
  in the simplest case of a structure  with
  a metallic gate and at high temperature.
 The Hall electric field is strongly suppressed  as
   compared with the case of one-component systems
  if the equilibrium  electron and hole densities
   are close  to each other
 and rapidly saturates to the result for a one-component system
 if the equilibrium  electron and hole
 densities becomes substantially different.

 We also studied the hydrodynamic
 corrections to the Hall effect and magnetoresistance
 in the ballistic regime, resulted from the
  arrival terms of the inter-particle collisions
  integrals.
  The hydrodynamic corrections, being
  a precursor of formation a viscous flow,
   arise due to electron-electron
   and hole-hole collisions conserving momentum  and protecting
a particle from the loss of its momentum at scattering on the sample
edges.

\section{  One-component systems }
\subsection{Model}
\begin{figure}[t!]
\centerline{\includegraphics[width=0.6\linewidth]{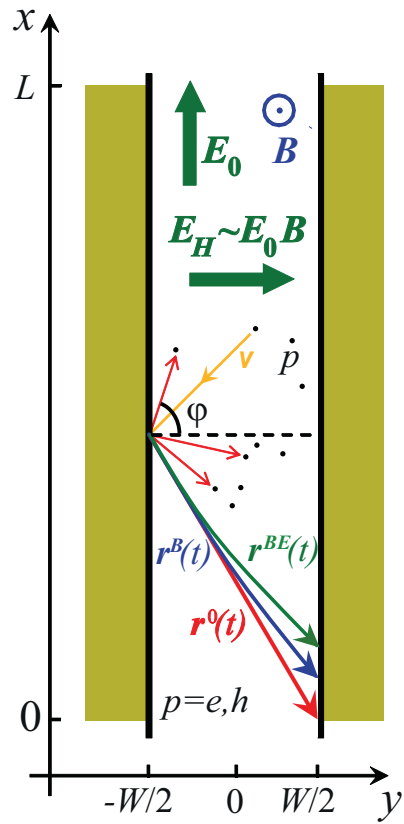}}
\caption{
A schematic illustration of a ballistic sample with particles $p$  and rough edges
 in external electric  and  magnetic fields $E _0 $ and  $B$.
 }\label{fig:1}
\end{figure}

 We consider  a flow of 2D charged particles
 (electrons or holes)
 in a long sample  with the width $W$
 and the length $L \gg W$ [see Fig.~\ref{fig:1}(a)]. We seek a linear response
 on a homogeneous generalized  external field $ {\bm E}_0||x $  in the presence of
  a magnetic field ${\bm B}$ perpendicular to the sample plane.
The amplitude of the field $ E_0 $ is proportional
 to a gradient of temperature for
 the problem of  heat transport and  coincides with an external
 electric field for the problem of charge transport.
If the mean free path relative to the inter-particle collisions, $l$,
 is much larger than the sample width $W$, $l\gg W$,
 the collisions with  the longitudinal  sample
 edges are the most frequent type of scattering
 events and the ballistic regime
  of heat or charged transport is realized.

In the current study we consider the sample to be enough clean and thus
 neglect particle scattering on disorder.
We assume the particle dispersion law to be quadratic.

The linear response of particles  to the external field $E_0$ is
described by the inequilibrium part of the distribution  function
 \begin{equation}
 \delta f(y,\mathbf{p}) = -f_F'(\varepsilon) f(y, \varphi,\varepsilon) \sim E_0
 \:,
 \end{equation}
where $f_F (\varepsilon)$  is the Fermi distribution function,
 $\varepsilon$ is the particle energy, $\varphi$
 is the angle of the particle velocity $\mathbf{v} = v(\varepsilon)
 [\, \cos \varphi , \, \sin \varphi \,  ]$,
 $\mathbf{p}=m \mathbf{v} $ is the particle momentum,
  and $m$ is the particle mass.
 see Fig.~\ref{fig:1}. The dependence of $\delta f$
  on the coordinate $x$
 is absent due to the relation $L \gg W$.

For simplicity, we use the rough approximation in which
 the energy dependencies
 in the absolute value of the particle velocity $v(\varepsilon)$
 and in  the factor  $f(y, \varphi, \varepsilon) $
 of the inequilibrium part of distribution function $\delta f(y,\mathbf{p})$
 are omitted.

  We use the system of units in which
  the characteristic particle
  velocity $v(\varepsilon) \equiv v$ is equal to unity and coordinates, time,
   and the reciprocal force from the generalized external field, $1/(eE_0)$,
   are measured in the same units. Here we introduce the dimensionless particle
   charge
   $e = \pm 1 $ in the expression for the external force $eE_0$
   in order to be able to specify the sign of the charge of
   particles for the problem of electric transport.

 The kinetic equation for the truncated distribution function
 $f(y , \varphi)  $ takes the form
 (see Ref.~\cite{ours_bal} and Fig.~\ref{fig:1}):
\begin{equation}
 \label{kin_eq}
\cos\varphi \, \frac{\partial f}{ \partial y }
 - \sin \varphi \, eE_0 - \cos \varphi \, eE_H +
  \omega_c \, \frac{\partial f}{ \partial \varphi } =
 \mathrm{St} [f]
 \:,
\end{equation}
where  the collision integral $\mathrm{St} $
describes the inter-particle scattering conserving
 momentum,  $\omega_c \sim eB $ is the cyclotron frequency, and
 $E_H$ is the Hall electric field arising
due to the presence of the magnetic field and
 related to redistribution of the 2D charged particles.
 In Eq.~(\ref{kin_eq}) we neglect scattering
  processes which do not conserve momentum.
 Following    Ref.~\cite{book__gas__the_method} and
  Refs.~\cite{hydr_tr_th_10,hydr_tr_th_10_2,hydr_tr_th_11,hydr_tr_th_11_2,
  hydr_tr_th_11_3},  we use the simplified form of the collision integral $\mathrm{St}$ :
\begin{equation}
 \label{St_N__St_U}
 \begin{array}{c}
  \displaystyle
\mathrm{St}  [f]  = - \gamma \, (f-P [f])
   \: ,
 \end{array}
\end{equation}
 where $\gamma$ is the
 scattering rate,  while  $P$ is
 the projector of distribution function $f(\varphi)$  on the subspaces consisting
 of the basis functions  $\{1,e^{\pm i \varphi} \}$.
 The operator $\mathrm{St}$ conserves the perturbations
 of the distribution function corresponding to a nonzero homogeneous flow
 and to a non-equilibrium concentration.

We assume the longitudinal sample edges being rough and
 the scattering of particles on them being fully diffusive.
Thus the boundary conditions on the distribution function  are as
follows:
 $ f(-W/2, \varphi)=c_l $ on the interval
 $ - \pi/2 < \varphi < \pi/2$  at the left sample edge, $y=-W/2$ (see Fig.~1),
   and $ f( W/2, \varphi) = c_r $
  on the interval  $  \pi/2 < \varphi < 3 \pi/2$
at the right edge. Here the quantities $c_{l} $ and $c_{r} $
   are the values of the distribution function $f$ averaged  over the angles of the particles
   trajectories
  reflected from the edges $y= \mp W/2$ \cite{rev_bal}:
  \begin{equation}
  \label{c_12}
  \begin{array}{c}
  \displaystyle
  c_l =  - \frac{1}{2} \int _{\pi/2} ^{3\pi/2}
   d \varphi' \: f(-W/2, \varphi') \cos\varphi'
     \: ,
   \\
   \\
   \displaystyle
   c_r = \frac{1}{2} \int _{-\pi/2} ^{\pi/2}
   d \varphi' \: f(W/2, \varphi')  \cos\varphi'
   \:.
   \end{array}
  \end{equation}
Such boundary conditions just express the fact that the $y$
component of the particle flow $j_y(y)$ vanishes at the edges (and thus
everywhere in the sample due to the continuity equation
$\mathrm{div}\,  \mathbf{j} =0$).

 The kinetic equation \eqref{kin_eq} can be rewritten as:
\begin{equation}
\label{kin_eq_with_gamma}
\Big[\cos\varphi \, \frac{\partial }{ \partial y }
  + \gamma \Big] \widetilde{f} - \sin\varphi \,eE_0
 = \gamma P [\widetilde{f} \, ] -
 \omega_c \, \frac{\partial \widetilde{f}}{ \partial \varphi }
  \:,
\end{equation}
 where we introduced the function
 \begin{equation}
 \label{f_tilda}
  \widetilde{f} (y,\varphi)= f(y,\varphi)  + e \phi(y) \:,
 \end{equation}
 in which $\phi$  is the electrostatic potential
 of the Hall electric field: $E_H = -\phi'$.

  In Ref.~\cite{ours_bal} we analyzed this kinetic equation in limiting regimes
  $\gamma W \gg 1$ and $\gamma W \ll 1$
  for the case of zero magnetic field, $\omega_c = 0$.
 We demonstrated that  in the hydrodynamic regime, $\gamma W \gg 1$,
  the left and the right parts of Eq.~(\ref{kin_eq_with_gamma})
 are of the same order of magnitude and Eq.~(\ref{kin_eq_with_gamma})
  transforms into the Navier-Stocks equation
 for the density of the particle (or heat) flow $j(y)  \equiv j_x(y)$:
 \begin{equation}
 \label{def_j}
  j(y) = \frac{n_0}{\pi m} \int _{0} ^{2 \pi } d \varphi \: \sin\varphi  \, f(y ,\varphi )
 \:,
 \end{equation}
  while  in the ballistic regime, $\gamma W \ll 1$,
 each term in the left part of \eqref{kin_eq_with_gamma}
 is   much larger than the right part term $\gamma P [\widetilde{f}]$.
  In Eq.~(\ref{def_j}) $n_0$ is the equilibrium density of particles
 (or particle energy)
and $m$ is the particle mass.
 Note that the exact form of Eq.~(\ref{def_j}) is due
 to the quadratic energy spectrum of the particles.

 For the case of the
 a nonzero magnetic field,
one can again prove that the arrival term $ \gamma P [\widetilde{f} \, ] $
in the ballistic regime, $\gamma W \ll 1$,
can be treated as a perturbation if  the magnetic field
is enough small [the  term
  $\omega_c \, \partial \widetilde{f}/  \partial \varphi $
  is smaller than  all  other terms in the left part of
 Eq.~(\ref{kin_eq_with_gamma})].
    In particular, this is the true for  the first-order by $\omega_c$ contribution
    to the particle distribution function $f_1 \sim \omega_c$, which describes the Hall effect.

For brevity,
further we will omit the tilde in designation of $ \widetilde{f} $
and just imply $f \equiv  \widetilde{f}$.

\subsection{Transport in zero magnetic field}
The solution of the kinetic equation (\ref{kin_eq_with_gamma}) with
the zero right part and the boundary conditions (\ref{c_12})  has the form
\cite{ours_bal}:
 $ f(y,\varphi) =   f_+(y,\varphi)$ at $  -\pi/2 < \varphi < \pi/2 $
and $  f(y,\varphi) =   f_-(y,\varphi) $ at
$ \pi/2 < \varphi < 3 \pi/2 $, where
\begin{equation}
  \label{f_pm}
f_{\pm} (y,\varphi)
 =
 e
  E_0\,\frac{\sin\varphi  }{\gamma}
 \Big[
 1-\exp \Big(
 \displaystyle
 -\gamma  \, \frac{
 y \pm W/2  }{ \cos \varphi }
  \Big)
 \Big]
 \:.
\end{equation}
For a long sample,   $L \gg 1/\gamma$,  the flow density
corresponding  to Eq.~(\ref{f_pm}) at $\gamma W \ll  1$
in the main order by the logarithm $\ln\left[1/(\gamma W\right)]$
is homogeneous:
\begin{equation}
 \label{j_0}
 j(y) = j _0   = \frac{2 e n_0 E_0 W }{ \pi m}  \ln\Big( \frac{1}{\gamma W }\Big)
 \:.
 \end{equation}
  For the total electric  current
\begin{equation}
  \label{def_I}
 I _0  =  e  \int   _{-W/2} ^{W/2} dy  \, j(y)
  \end{equation}
  we  obtain:
\begin{equation}
  \label{I_0}
I _ 0  = \frac{ 2 e^2 n_0 E_0  W^2 }{ \pi m}\ln\Big(\frac{1}{\gamma W}\Big)
\:.
\end{equation}

 It is seen from Eq.~(\ref{f_pm}) that the
 logarithmic divergence  in  $j_0$ is related to the particles  with the
 velocity angles $\varphi $ in the diapason
 $||\varphi | - \pi/2 | \lesssim \delta_m $, where
  $\delta_m  = \gamma W \ll  1$ is
   the characteristic
 value of the difference $|\pi/2- |\varphi||$
 corresponding to the particles giving the main contribution to the current.
  Such  particles are moving  almost parallel to the sample direction.
  A particle
 on such ``special'' trajectories  spends a longer time
between scattering events  on the opposite edges as compared to the
particles moving   along the ``regular'' trajectories  with $\varphi
\sim 1 $ and,  thus, acquires a larger  velocity correction due to
acceleration by the field $ E _0 $.

A more exact solution of
 Eq.~(\ref{kin_eq_with_gamma})
 with taking into account the arrival term in the collision integral, $\gamma P[f]$,
 provides a hydrodynamic correction
 $\delta I _h$ to the current $I_0$ \cite{ours_bal}. Such correction is related to
 the inter-particle collisions conserving momentum,
 which protect particles from a loss of their momentum in
  scattering on  edges.
In order to calculate the hydrodynamic
correction in the ballistic limit, $\gamma W \ll 1$,
we present the distribution function  in the form
$ f=f_0+f_1 $,  where $f_0$ is the function (\ref{f_pm})
 and $f_1$ is a correction to $f_0$  corresponding to the non-zero  right part
  of Eq.~(\ref{kin_eq_with_gamma}), $\gamma P[f]$.
 The equation for $f_1$ takes the form:
\begin{equation}
 \label{Eq_f0_f1}
   \Big[\cos\varphi \, \frac{\partial }{ \partial y } + \gamma \Big] f_1
 = \gamma   P_{\sin} [f_0]
 \:,
\end{equation}
where $P_{\sin}$ is the projector on the function $\sin \varphi$.

Action of the operator $P_{\sin}$  on the zero-order distribution function
 $f_0$ yields a value proportional to the current density:
    $P_{\sin}[ f_0] = j(y) \sin \varphi / (n_0/m)  $,
  where $j(y) = j_0 $ is given by Eq.~(\ref{j_0}). Therefore,  the right part
of Eq.~(\ref{Eq_f0_f1})  becomes equal just to $ q E_0 \sin \varphi $,
    where
 \begin{equation}
  q= \frac{ 2 }{\pi} \, \gamma  W   \, \ln \Big( \frac{1}{\gamma  W} \Big)   \ll 1
  \:.
 \end{equation}
  By this way, Eq.~(\ref{Eq_f0_f1}) turns  into
  Eq.~(\ref{kin_eq_with_gamma})  with zero right part
  and the value $E_0$ replaced by $qE_0$. Thus for all
  the values related to the  first hydrodynamic
  correction $f_1$  we just have in the main order
  by the logarithm $\ln[1/(\gamma  W)]$:
  $ f_1 (y, \varphi) = qf_0 (y, \varphi) $,
  $ j_1 (y) = q j_0 (y) $,
   and $ I_1  \equiv \delta I_h = q I_0 $, namely:
\begin{equation}
  \label{delta_I_H}
   \delta I _h = \frac{4 e^2  n_0 E_0  \gamma  W^3}{\pi^2 m }
    \ln\Big(\frac{1}{\gamma W}\Big)^2
 \:.
\end{equation}
 The obtained correction \eqref{delta_I_H} is positive and   originates from the small group of
    particles whose last scattering event was an inter-particle collision.
    By this way, arising of the correction
     $\delta I_h$ due to such particles   is a {\em precursor } of forming
the Poiseuille flow of a viscous fluid
 related to the inter-particle collisions conserving momentum.

\subsection{Magnetotransport }
\subsubsection{The Hall effect }
In this subsection we study the Hall effect and magnetoresistance
of the one-component system in the ballistic regime, $\gamma W \ll 1$,
 within the kinetic equation (\ref{kin_eq_with_gamma}).

As it was discussed above, at enough small $\omega_c$
the arrival term $\gamma  P[f]$
 in the main order  by the logarithm $ \ln[1/(\gamma W) ]  \gg 1$ can
 be neglected in the main
  approximation by $\gamma$ and the kinetic equation (\ref{kin_eq_with_gamma}) takes the form:
\begin{equation}
\label{kin_eq_B_main}
\left[\cos\varphi \,  \frac{\partial }{ \partial y }
  + \gamma \right] f
+  \sin \varphi \,e E_0
 =  - \omega_c \frac{\partial f}{ \partial \varphi }
 \:.
\end{equation}
 The right part of this equation  is a small perturbation as compared to the left part.

We seek the solution of Eq.~(\ref{kin_eq_B_main})
 in the form of the series $f=f_0+f_1+f_2$,
 where  $f_0$ is given
 by Eq.~(\ref{f_pm}), while $f_1$ and $ f_2$ are proportional
 to the powers of magnetic field:
 $ f_1\sim \omega_c\sim B $,  $ f_2 \sim  \omega_c^2\sim B^2 $.
   For the functions $f_1$ and $f_2$
  we obtain from Eq.~(\ref{kin_eq_B_main}):
 \begin{equation}
 \label{B_f_1}
 \Big[\cos \varphi  \frac{\partial }{ \partial y } + \gamma \Big] f_1
  = - \omega_c \frac{\partial f_0 }{ \partial \varphi }
 \: ,
\end{equation}
 \begin{equation}
 \label{B_f_2}
 \Big[\cos \varphi   \frac{\partial }{ \partial y } + \gamma \Big] f_2
  =- \omega_c \frac{\partial f_1 }{ \partial \varphi }
  \:.
\end{equation}

 The solution of Eq.~(\ref{B_f_1}) for the zero boundary
  conditions with $c_{l,r}\equiv 0$ in Eqs.~(\ref{c_12})
 is
  \begin{equation}
 \label{f_1}
 \begin{array}{c}
 \displaystyle
f_1 ^z  (y,\varphi) = - \omega_c e E_0
 \Big\{
 \frac{\cos \varphi }{\gamma^2 }
 - \exp \Big[- \gamma \,   \frac{y  \pm  W/2  }{ \cos \varphi }\Big]
\\
\\
\displaystyle  \times
 \Big[ \frac{\cos \varphi }{\gamma^2 }
  + \frac{ y \pm W/2 }{\gamma} -
  \frac{ \sin ^2 \varphi }{ 2 \cos ^3 \varphi
} \, \Big(y  \pm \frac{W}{2} \, \Big)   ^2
 \,  \Big] \, \Big\} ,
\end{array}
 \end{equation}
 where the signs $\pm$ corresponds to the diapasons
 of the angles $-\pi/2 < \varphi < \pi/2
 $  and $ \pi/2 < \varphi < 3\pi/2
 $,  respectively.
It can be seen from comparison of   Eqs.~(\ref{f_pm}) and
(\ref{f_1}) that the perturbation theory by the magnetic field term
 can
be used in low magnetic field  until
 \begin{equation}
 \omega_c \ll \gamma^2W
 \:.
 \end{equation}

 The function $f_1$  satisfying the non-zero boundary conditions
    with $c_{l,r} $ from Eqs.~(\ref{c_12})
    has the form: $f_1=f_1^z+\delta f_1$,
    where
    \begin{equation}
    \label{delta_f_1}
    \delta f_1 ( y , \varphi) = C _{\pm}
    \exp \Big(  -\gamma \frac{ y \pm W/2 }{\cos \varphi} \Big)
    \end{equation}
     is some  solution of Eq.~(\ref{B_f_1})
    with the zero right  part.
    A straightforward calculations based on (\ref{f_1})
 lead to the proper values of the coefficients $C_{\pm} $:
 \begin{equation}
 \label{C}
  C_{\pm} = \pm \omega_c e E_0 \frac{W}{4\gamma}
  \:.
  \end{equation}
  The resulting correction
       $ \delta f _1 $
      is much smaller than $f_1^z$ at the angles
      $||\varphi | -  \pi/2 | \lesssim \delta_m $.

If a current flows through a sample
in magnetic field, a perturbation of
the charged particle density and
 the Hall electric field arises due to
the magnetic Lorentz force.
 In our system these effects are described
by the zero angular harmonic  of the function $f_1$
\begin{equation}
\label{zero_harm_def}
 f^{m=0}_1 (y) = \frac{1}{2\pi} \int _0 ^{2\pi}
d\varphi
 \:
f_1(y,\varphi)
 \:.
\end{equation}
From Eqs.~(\ref{f_1}),
 (\ref{delta_f_1}), (\ref{C}), and (\ref{zero_harm_def})
 in the main order by $1/(\gamma W)$
  and $\ln[1/(\gamma W)]$ we obtain:
\begin{equation}
 \label{f_1_m_0}
 f_1^{m=0} (y) =
  -\frac{\omega_c e E_0  }{\pi} \, W y
  \, \ln\Big(\frac{1}{\gamma W}\Big)
 \end{equation}
 The other terms in decomposition of $f_1^{m=0}$
  by the parameter $ \gamma  W$
 are of the smaller orders of magnitude:
 they are proportional to
   $ \omega_c eE_0W^2 (\gamma  W)^k $, where $k \geq 0$.

The zero harmonic of the distribution function (\ref{f_tilda}) is
$\delta \mu (y) + e \phi(y)$, where $\delta \mu$ is the perturbation
  of the particle chemical  potential.
 For the gated as well as the ungated
 structures with a one-component 2D system
  of charged particles,
 the electrostatic potential $\phi$ is usually much
  greater than the the perturbation of the chemical potential
 $\delta \mu$
  (see, for example, Ref.~\cite{visc_res}).
  As a result,  Eq.~(\ref{f_1_m_0}) yields the expression for the
 Hall electric field $E_H = const(y)$:
 \begin{equation}
 \label{E_H_main}
 E_H  = \frac{ \omega_c E_0 }{\pi} \, W
 \, \ln\Big(\frac{1}{\gamma W}\Big)
 \:.
 \end{equation}

Note that the result (\ref{f_1_m_0})
 for the zero harmonic of $f$  was calculated
 from the kinetic equation in the form (\ref{kin_eq_B_main})
  taking takes into account only the departure term
  in the inter-particle collision integral and the magnetic field term.
 Such equation describes the particles which,
 after scattering on one sample edges, reach the other edge
 or go out of consideration [within Eq.~(\ref{kin_eq_B_main})]
  due to inter-particle scattering at $-W/2<y<W/2$.
Thus  the Hall effect
 in the ballistic regime is due to
 an inhomogeneous distribution of the
  particles densities
 resulted from their collisionless motion in the external fields and
 scattering-related departure out of consideration.
    In contrast to the Ohmic and the hydrodynamic regimes,
  the Hall electric field in the ballistic regime
   do not compensate the magnetic force acting on some
  small fluid elements with a quasi-equilibrium  distribution of particles.

According to the kinematic nature of the Hall effect,
the Hall field (\ref{E_H_main})
depends on the scattering rate $\gamma$ only via
 the logarithm describing the characteristic minimal value of
$ | |\varphi| - \pi/2 |$.

From comparison of Eq.~(\ref{j_0}) and (\ref{E_H_main})
one can calculate the Hall coefficient $R_H = E_H/(B j)$:
 \begin{equation}
 \label{R_H}
 R_H  = \frac{1}{2}R_H ^0
  \:, \qquad
   R_H ^0 = \frac{1}{n_0 e c }
 \:,
 \end{equation}
where $R_H ^0$ is the conventional Hall coefficient for
 a quasi-equilibrium (Ohmic of hydrodynamic) flow of particle
  with the quadratic spectrum at  low temperatures.

In the recent experimental work \cite{Gusev_3}
 the Hall resistance of narrow samples of ultra-high-mobility
 GaAs quantum wells  was measured.
 A deviation of the Hall coefficient
 from its ``quasi-equilibrium'' value $R_H ^0$
 was observed in weak and moderate magnetic field
 when the giant negative magnetoresistance is observed in such structure.
 The Hall coefficient near the zero magnetic field turned out
  to be 20 percent less than $R_H ^0$,
 and with the growth of the magnetic field it
 becomes   larger than $R_H ^0$,
  and then it become very close to $R_H ^0$.
   The observed value of the Hall coefficient, $R_H ^{exp} < R_H ^0$,
   at the very small magnetic field
  qualitatively agrees with the result~(\ref{R_H}).
Such behavior of the Hall field  and  magnetoresistance corresponds
  to the crossover from the ballistic to the hydrodynamic regimes
   of transport taking place
   in narrow ($\gamma W  \ll 1 $) samples
   with the increase of magnetic field.

\subsubsection{Magnetoresistance}
Substitution of Eq.~(\ref{f_1}) to
 Eq.~(\ref{B_f_1})  and solving the resulting equation
 with the zero boundary conditions yields
  the correction $f_2$ to the distribution function
   of the second order by magnetic field.
  For the angles $||\varphi| - \pi/2| \ll 1$ in the main order
  by $1/(\gamma W) $ it can be written as \cite{ours_bal}:
 \begin{equation}
 \label{f_2}
\begin{array}{c}
\displaystyle
f_2^{ \pm}(y,\delta) =
 \frac{\omega_c ^2  E }{ 2\delta^5 } \Big(y \pm \frac{W}{2} \Big) ^3
\times
\\
\\
\displaystyle
\times \Big[ \, 1 -\frac{1}{4} \Big(y \pm \frac{W}{2} \Big)
 \frac{\gamma}{\delta} \, \Big]
 \:  \exp\Big[ - \frac{\gamma}{\delta}
  \Big(y \pm \frac{W}{2} \Big) \Big]
 \:,
 \end{array}
 \end{equation}
 where $\delta = \cos \varphi$.
This contribution to the distribution function results in the
correction to the total current $I$ of the second order
   by the magnetic field \cite{ours_bal}:
  \begin{equation}
  \label{final}
 I = I_0 + \Delta I
 \,,\;\;\;
 \Delta I = \frac{3 e^2 n _0 E_0  }{2 \pi m}
  \frac{ \omega_c^2  }{  \gamma^4 }
  \:.
 \end{equation}
This result corresponds
 to a small negative magnetoresistance.
  For  long samples, $L \gg 1/\gamma$,
Eqs.~(\ref{final}) lead to \cite{ours_bal}:
 \begin{equation}
  \label{MR_B_straight_gamma}
 \frac{R (B) - R(0) }{  R(0)}  =  - \frac{ 3 \omega_c^2   }
 { \displaystyle 4 \gamma^4 W^2 \ln\Big(\frac{1}{\gamma W } \Big)}
 \:, \;\;\;
 \omega_c \ll \gamma ^2 W
  \:,
  \end{equation}
where $R(B) = EL/I(B)$ is the sample resistance. The physical origin
of the obtained magnetoresistance is in an increase
 of the mean length of the electron trajectories
 by which electrons move from one edge to another
 without inter-particle collisions
  (see Fig.~1 in Ref.~\cite{ours_bal}).

\subsubsection{Hydrodynamic corrections }
The corrections to the distribution functions
 (\ref{f_1}) and (\ref{f_2})
  due to the arrival term $\gamma P [f] $ of the collision
  integral $\mathrm{St}$ leads to  the  next orders
 contributions by  the parameter $\gamma W$ to the Hall electric field
 and magnetoresistance.

In order to calculate such the hydrodynamic  corrections,
the kinetic equation (\ref{kin_eq_with_gamma})
 should be solved by the  perturbation
theory by the both terms
$\gamma P [\widetilde{f} \, ] $ and
$ \omega_c \, \widetilde{f}/  \partial \varphi $.
 A straightforward analysis shows that these corrections are
 similar by their origin and structure
 to the hydrodynamic correction (\ref{delta_I_H})
 to the current in zero magnetic field.
In the first order by  $\gamma W$ and in the main order
 by the logarithm $\ln[1/(\gamma W)]$ we obtain:
\begin{equation}
\label{corr_fist}
 \Delta E_H  = q \,  E_H
  \, , \qquad
  \Delta I _2 =  2 q\, \Delta I
  \:,
\end{equation}
that leads to:
\begin{equation}
\label{corr_to_E_H}
 \Delta E_H
 = \frac{2 \omega_c \gamma W^2 E_0}{\pi^2} \,
 \ln ^2 \Big(  \frac{1}{\gamma W} \Big)
 \end{equation}
and
\begin{equation}
\label{corr_to_MR}
 \frac{ \Delta R (B)  }{  R(0)}  =  - \frac{ 3
\omega_c^2   }
 {  \pi  \gamma^3 W }
\:.
\end{equation}

It is noteworthy that the hydrodynamic correction~(\ref{corr_to_E_H}) to the Hall electric field has the same sign as the main ballistic contribution~(\ref{E_H_main}). This indicated that inter-particle collisions, leading to formation of a viscous flow, induce the increase of the ballistic Hall coefficient $R_H=1/(2n_0ec)$ [see Eqs.~(\ref{R_H})], approaching it to the Hall coefficient in the hydrodynamic regime $R_H^0=1/(n_0ec)$.

\subsubsection{The case of not very long samples}
 For the samples with the widths in the interval  $W  \ll L \ll 1/\gamma$
 the parameter $\delta _m  $, characterizing
 the  typical values of the difference $| \pi/2- |\varphi||$
 for the most important particles,
  can be estimated as $\delta_ m  \sim W/ L $.
    In this case, the scattering of particles in the contacts
    located at $x=\pm L/2$ is more frequent than inter-particle collisions.
    The rate of such scattering in  contacts can be estimated as $1/L$.
    Therefore in order to estimate the current, magnetoresistance and the Hall effect,
    one can apply the formulas already obtained for long samples,
    $L \gg 1/\gamma$, replacing in them $\gamma$ on $1/L$.
    Herewith  both the departure  and the arrival terms
    in the collision integral of inter-particle
    will play the role of additional small perturbations.

  According to such procedure, first,
  we obtain the result  for the total current $I_0$.
 In the main order by the logarithm $\ln(L/W)$ we have \cite{ours_bal}:
\begin{equation}
  \label{I_0_L}
I _0 =   \frac{ 2e^2 n_0 E_0  W^2 }{ \pi m}\ln \Big( \frac{L}{W} \Big).
 \end{equation}
 Second,
  the expression  (\ref{E_H_main})
   for the Hall electric field
   changes to \cite{n}:
    \begin{equation}
 \label{E_H_main_not_very_long}
 E_H  = \frac{ \omega_c E_0  W}{\pi}  \, \ln\Big(\frac{L}{W}\Big)
 \:.
 \end{equation}
For magnetoresistance   instead of  Eq.~(\ref{MR_B_straight_gamma})
we obtain \cite{n}:
 \begin{equation}
  \label{MR_B_straight_L}
 \frac{R (B) - R(0) }{  R(0)}  \sim
  - \frac{ \omega_c^2 L^4  }{ W^2 \ln(L/W )}
 \:,
 \;\;\; \omega_c \ll \frac{W}{L^2}
 \:.
\end{equation}
It is noteworthy that the Hall electric field and magnetoresistance in
this approximation are independent of $\gamma$ and, thus, of temperature.

 The temperature dependencies of the current,
 magnetoresistance, and the Hall effect
 are controlled by the the collision integral $\mathrm{St}[f]$
 being a small perturbation in a whole
 (both the departure as well as arrival terms).
The corrections from the departure term $-\gamma f$
can be obtained from Eqs.~(\ref{corr_to_E_H}) and
(\ref{corr_to_MR}) just by replacement
 $1/L \to 1/L + \gamma $, where $1/L \gg \gamma$.
 Eqs.~(\ref{corr_fist}) yield the corrections from the arrival term
 $\gamma P[f]$:
\begin{equation}
\label{corr_to_E_H_not_long}
 \Delta E_H
 = \frac{ 2 \omega_c   \gamma W^2 E_0}{\pi^2}  \, \ln^2 \Big(\frac{L}{W}\Big)
 \end{equation}
and
\begin{equation}
\label{corr_to_MR_not_long}
 \frac{\Delta R (B) }{  R(0)}
  \sim
  - \frac{ \omega_c^2 \gamma L^4  }{ W }
\:.
\end{equation}
It can be seen from
Eqs.~(\ref{E_H_main_not_very_long})-(\ref{corr_to_MR_not_long})
 that the last corrections from the arrival term
  dominates on the corrections from the  the departure term
 in the main order by the logarithm $\ln[1/(\gamma W )] \gg 1$.

\section{ Two-component systems }
\subsection{Model}
We consider a two-component two-dimensional electron-hole
system
 in a sample on a substrate with a metallic gate. In such setup,
the equilibrium particle densities
 can be controlled by varying the gate voltage. We choose
 a simplest model for the two-component
 system: the electron and hole bands are separated
 by the gap $\Delta$, their energy spectrums are quadratic with the same effective masses
 and being fully symmetric relative to the point $\varepsilon = 0$,
 and  the  chemical potential $\mu_0$ lies inside the gap near the point
 $\varepsilon = 0$ [see Fig.~2].

\begin{figure}[t!]
\centerline{\includegraphics[width=0.5\linewidth]{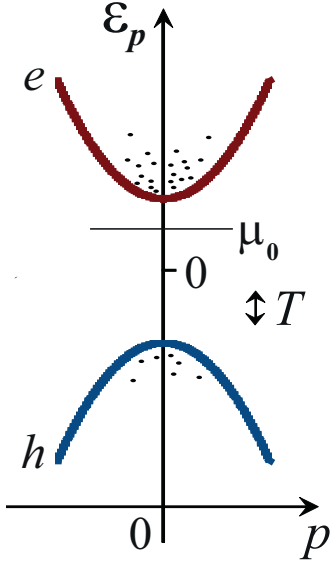}}
\caption{
 A band diagram of the simplest two-component
 electron-hole system with the symmetric  quadratic dispersion laws and the
non-degenerate statistics.
 }\label{fig:2}
\end{figure}

If $\mu_0, T \ll \Delta $, the particles of the both types form
 non-degenerate Boltzmann gases. For simplicity, we will use the formulas
 for the non-degenerate gase up to the values of the chemical
 potential $\mu_0 ,T \lesssim \Delta /2$, when some corrections
 to the all formulas due to the Pauli principle arise.
 Such an approximation corresponds in its accuracy to the neglect, made in the previous section,
  of the dependencies
  of the function $f$ and  the particle velocity $v$ on the particle energy $\varepsilon$.

 We again consider a flow in a long  sample
 with the length $L$ and the width $W\ll  L $.
 The sample edges is supposed to be rough and
 the scattering of particle on the edges is fully diffusive.
   In this section we present the results only
 for the  very long samples, $L \gg 1/ \gamma_m$, where $\gamma_m$
 is the minimum value of the scattering rates in the system.

 The kinetic equation for a two-component system
 in the simplest approximation used in the previous section
 for a one-component system
 takes the form:
 \begin{equation}
 \label{kin_eq___two-c}
 \begin{array}{c}
 \displaystyle
\cos\varphi \, \frac{\partial f^{\alpha}}{ \partial y }
 - \sin \varphi \, e^{\alpha} E_0 - \cos \varphi \, e^{\alpha} E_H +
 \\
 \\ \displaystyle
  \qquad \qquad
  +
  \omega_c^{\alpha} \, \frac{\partial f^{\alpha}}{ \partial \varphi }
  =
 \mathrm{St}^{\alpha\alpha}  [ \, f^{\alpha} \, ] +
 \mathrm{St}^{\alpha\overline{\alpha}}
  [ \, f^{\alpha},f^{\overline{\alpha}} \, \,]
  \:,
 \end{array}
 \end{equation}
 where the superscript $\alpha = h,e$ denotes the type of particles:
 $\alpha = h$ for holes and $\alpha = e$ for electrons;
 $\overline{\alpha} = e$ for $\alpha = h$ and
$\overline{\alpha} = h$ for $\alpha = e$; $e^{h,e} =\pm e_0 $, $e_0$
is the absolute value of the electron charge;
 $\omega_c^{\alpha} =e^{\alpha}B/mc$.
 The collision integrals
 \begin{equation}
 \mathrm{St}^{\alpha\alpha} [f^{\alpha}]=
 \gamma^{\alpha} (f^{\alpha} -P [f^{\alpha} \, ] )
 \end{equation}
describe electron-electron and hole-hole collisions,
both conserving momentum
 and leading to forming hydrodynamic electron and hole flows
 (if only one component of the electron-hole fluid is present).
 The collision integrals
 \begin{equation}
  \mathrm{St}^{\alpha\overline{\alpha}} \,
   [ \, f^{\alpha},f^{\overline{\alpha}} \, \,]=
 \gamma ^ { \alpha\overline{\alpha}} (f^{\alpha}
 -P [f^{\overline{\alpha}}\, \, ] \,  )
 \end{equation}
 describe electron-hole collisions,
 which also conserve  the total momentum,
 but lead to a finite Ohmic resistance
 for an electron-hole system in electric field
 (due to opposite direction of the electric forces
 acting on electrons and on holes).
  For the relaxation rates in a symmetric non-degenerated electron-hole
  system shown at Fig.~2
 we have: $\gamma^{\alpha} = \Gamma n_0 ^{\alpha} $
 and $\gamma^{\alpha\overline{\alpha} } = \Gamma n_0 ^{\overline{\alpha}} $,
 where  $n_0 ^{h,e}  = T \nu e^{-\Delta /(2T) \mp \mu_0/T} $
 are the equilibrium particle
 densities,
 $\Gamma$ is a parameter independent on the densities $n_0 ^{\alpha}$,
  $\nu = g m / (2\pi \hbar^2)$ is the 2D density of states
  for the particles with a quadratic spectrum,
  and $g$ is the spin-valley degeneracy.

As it has been done
 for a one-component system, the kinetic
equation (\ref{kin_eq___two-c})  can be rewritten in the form:
\begin{equation}
\label{kin_eq_with_gamma___two_c}
\begin{array}{c}
\displaystyle
\Big[\cos\varphi \, \frac{\partial }{ \partial y }
  + \gamma \Big] \widetilde{f}^{\alpha} - \sin\varphi \,e^{\alpha} E_0
 =
 \\
 \\
 \displaystyle \qquad \qquad
 =
 \gamma^{\alpha} P [\widetilde{f} ^{\alpha} ] +
 \gamma^{\alpha \overline{\alpha} } P [\, \widetilde{f} ^{\, \overline{\alpha}} \, ]
  - \omega_c^{\alpha} \, \frac{\partial \widetilde{f} ^{\alpha} }{ \partial \varphi }
  \end{array}
  \:,
\end{equation}
where the terms in the right part should be considered as small
perturbations to the left part. In
Eq.~(\ref{kin_eq_with_gamma___two_c}) we introduced the notations
$\widetilde{f}^{\alpha}  = f^{\alpha} +e^{\alpha}  \phi $ and
$\gamma = \gamma^{\alpha} + \gamma^{\alpha\overline{\alpha} }  = \Gamma (n_0^e + n_0^h )$.
For brevity, we again omit the tilde in the functions $\widetilde{f}^{\alpha}$
and further write just $f^{\alpha} \equiv \widetilde{f}^{\alpha}$.

\subsection{Magnetotransport and electrostatics}
We see that the left part of Eq.~(\ref{kin_eq_with_gamma___two_c})
 and the magnetic field term in its right part  are
  identical with the ones of Eq.~(\ref{kin_eq_with_gamma}).
  Thus the expressions
 for the main parts of the distribution function
 $f_0$ and the current density $j_0$ as well as for
 the magnetic field corrections $f_1 \sim \omega_c$ and  $f_2 \sim \omega_c^2$
 obtained in the previous section
  for a one-component systems remains valid for each of
  the both electron and hole  components of the two-component system.
  In this way, we have:
  \begin{equation}
\label{j_0_two_com}
 j_0^{\alpha} =  \frac{2 n ^{\alpha}  _0 e ^{\alpha} E_0  W  }{\pi m}
  \, \ln\Big( \frac{1}{\gamma W }\Big)
  \:,
\end{equation}
   \begin{equation}
\label{I_0__two_com}
 I_0  = \frac{ 2 e_0^2 (n_0^h + n_0^e)  E_0  W^2 }{\pi m}
 \, \ln\Big(\frac{1}{\gamma W}\Big)
 \:,
\end{equation}
   \begin{equation}
\label{Delta_I__two_com}
 \Delta I   =  \frac{ 3   e _0 ^2   ( n^ h + n^e_0) E_0    }{2 \pi  m}
  \frac{ \omega_c ^2  }{  \gamma^4 }
  \:,
\end{equation}
  and
  \begin{equation}
\label{f_1__two_res}
  (f_1 ^{\alpha} )^{m=0}  = - \frac{\omega_c e _0  E_0 }{ \pi}
 \, W y  \, \ln\Big(\frac{1}{\gamma W}\Big)
 \:,
\end{equation}
  where $\omega_c = |\omega_c^{\alpha}|$.
   We see from Eq.~(\ref{j_0_two_com}) that the hole and the electron
 flows have opposite directions: $j_0^{h,e} \sim \pm n^{h,e}_0 $.
The formulas (\ref{I_0__two_com}) and (\ref{Delta_I__two_com}) leads
to the negative magnetoresistance identical to the magnetoresistance
(\ref{MR_B_straight_gamma}) in a one-component system.

If  $\mu_0 = 0 $  the electron-hole system becomes fully symmetric
 relative to the operation of replacing of electrons by holes and changing
 the directions of all the fields. In particular,
   the charge density is equal to zero as $n_0^e = n_0^h $ and
   $\delta n^e = \delta  n^h $.
 This is the so-called charge-neutrality point.
 In it
   the Hall electric field is absent,  and
   the zero harmonic of the magnetic field correction to the
 distribution function $f_1$ is related only to a perturbation of the
 total particle density $ \delta n^e  + \delta n^h  $.

At the gate voltages corresponding to small values of the
equilibrium chemical potential,
 $\mu_0 \ll T , \Delta  $, the electron and the hole components of the system
 as well as their responses on the external fields are almost symmetrical.
 In particular, the electrons and the hole
 equilibrium densities  $n_0^e$ and $n_0^h$ as well as their perturbations
 $\delta n^e $ and $\delta n^h$
 are close one to another: $ | n_0^e - n_0^h | \ll n_0^{e,h} $
 and  $| \delta n^e -\delta n^h| \ll |\delta n^{e,h}| $.
 In such situation, the zero harmonic $(f_1^{\alpha})^{m=0}$
 of the distribution function
 is related to both  a perturbation of the
 particle density $\delta  \varrho   $ and
  arising of a charge density $e_0(\delta n^h -\delta n^e ) $
 (related to the Hall electric field):
    \begin{equation}
    \label{f_1__two_def}
    (f_1 ^{\alpha} )^{m=0} = \delta \mu ^{\alpha }  + e ^{\alpha} \phi
    \:,
    \end{equation}
    where $\delta \mu ^{\alpha } $ are the perturbations of the electron
    and the hole chemical potentials and
    $\phi$ is the electrostatic potential corresponding
    to the charge density $e_0(\delta n^h -\delta n^e ) $.
Using Eqs.~(\ref{f_1__two_def}) and (\ref{f_1__two_res})  one can  calculate
     the perturbations of the particle densities
     \begin{equation}
     \delta n ^{h,e}= \nu \, \delta \mu  ^{h,e}
      \exp \Big( \displaystyle - \frac{\Delta }{2T} \mp \frac{\mu_0 }{T} \Big)
     \end{equation}
     and the Hall electric field $E_H  = - \phi' $
     in a vicinity as well as far from the charge neutrality point.

\begin{figure}[t!]
\centerline{\includegraphics[width=0.95\linewidth]{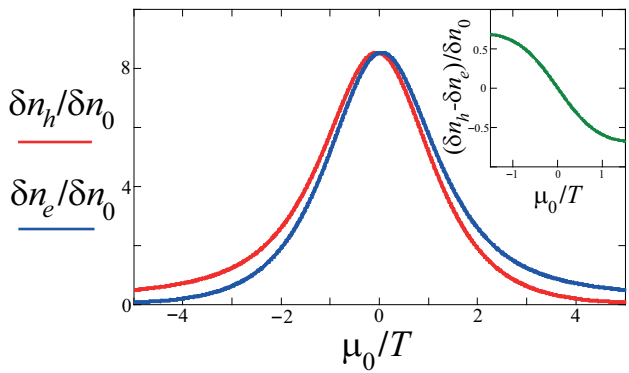}}
\caption{
  The dependence of the perturbations of the hole and
  electron  densities $\delta n_h $ and $\delta n_e $
  at the sample edge $y = - W/2  $
   on  the equilibrium chemical potential $\mu_0$. Inset shows
   the difference $(\delta n_h - \delta n_e)|_{y = - W/2  }/ \delta n_0 $
   which is proportional to the charge density and the Hall electric field $E_H$.
   All curves are plotted for the following parameters:
   $\gamma |_{\mu_0=0} W =  2 \cdot 10^{ -4}$ and  $ D= 10 $.
  }
\end{figure}

We describe the electrostatics of the two-component system
 and the metallic gate within the gradual channel approximation,
 in which the electrostatic potential
 $\phi$ is related to the charge density in the 2D layer
 as:
 \begin{equation}
 \label{electrostatics}
 \phi (y)= \frac{ 4 \pi e_0 d}{\kappa }
  \, [\,  \delta n ^h(y) -\delta n ^e
 (y ) \,]
 \:.
 \end{equation}
 Here  $\kappa$ is the background dielectric constant and
 $d$ is the distance between the 2D layer and the gate,
 which should be much greater than
 the Bohr $a_B$ and the Debye $r_D$ radii  of 2D particles.
 Solving the system of equations~(\ref{f_1__two_res}),
  (\ref{f_1__two_def}), and (\ref{electrostatics})
  for $\delta \mu ^e $, $\delta \mu ^h $ and $\phi$,  we obtain:
 \begin{equation}
 \label{delta_n_res}
 \frac{\delta n ^{h,e} (y,\mu_0)}{\delta n _0}
 =
 -
  \frac{D + e^{ \mp \mu_0 /T }/2}{ D \cosh (\mu_0/T)}
 \, \frac{y}{W} \,  \ln \Big(  \frac{1}{\gamma W } \Big)
 \:,
  \end{equation}
where the parameter  $D$,
\begin{equation}
D = \frac{2g   d}{a_B }\exp \Big( \displaystyle  - \frac{\Delta }{2T} \Big)
\end{equation}
  is
considered to be much greater than unity (it is possible as $d \gg
a_B$) and the amplitude
  \begin{equation}
  \delta n _0 = \frac{\omega_c e_0 E_0 W ^2 }{\pi}
  \, \, \nu \,   \exp\Big(\displaystyle  - \frac{\Delta }{2T}\Big)
   \end{equation}
does not depend on $\mu_0$. For the Hall electric  field, $E_H =
-\phi' = const(y) $, we obtain from (\ref{electrostatics})
 and  (\ref{delta_n_res}):
 \begin{equation}
 \label{E_H_n_res}
E_H(\mu_0) = - \frac{ \omega_c  W  E_0 }{\pi}
 \, \, \tanh\Big( \frac{\mu_0 }{T } \Big)
 \, \ln \Big(  \frac{1}{\gamma W } \Big)
 \:.
  \end{equation}
In Fig.~3 we plotted the values $\delta n ^{h,e}$ at the sample edge
$y = - W/2$ as functions of the equlibrium chemical potential $\mu_0$
with taking into account the dependence of the rate $\gamma =
\gamma^{\alpha\alpha} + \gamma^{\alpha\overline{\alpha}}$ on the
equilibrium densities $n_0^{\alpha} = n_0^{\alpha}(\mu_0)$.
 We see that near the charge neutrality point, $\mu_0 = 0$,
the perturbation of the electron and hole densities are close one to another, while with
 the removal of  $\mu_0$ from the point
$\mu_0 = 0$ the electron and hole densities are very different and the Hall electric field (\ref{E_H_n_res}) approaches to the result (\ref{E_H_main}) obtained in the previous section for a one-component system.

\subsection{Hydrodynamic and Ohmic corrections}
From the kinetic equations in the form
(\ref{kin_eq_with_gamma___two_c}) one can calculate the hydrodynamic
and the Ohmic corrections to the current and the Hall electric field
related to the arrival part of collision integrals
$\mathrm{St} ^{\alpha\alpha}$ and $\mathrm{St}
^{\alpha\overline{\alpha}}$, respectively.
The terms ``hydrodynamic'' and ``Ohmic
corrections'' have following origins.
 Due to opposite
signs of the electron and the hole charges, $e^{e,h} = \mp e_0$,
 the particle flows $j_0^{e}$ and $j_0^{h}$ have opposite
directions [see Eq.~(\ref{j_0_two_com})]. Thus
 the corrections to the particle flows $j_0^{\alpha}$
  due to the arrival terms $\gamma ^{\alpha } P[f^{\alpha }]
  $ are co-directed with $j_0^{\alpha}$,
 while the corrections to $j_0^{\alpha}$
  from  the departure  terms $\gamma ^{\alpha \overline{\alpha } }
  P[f^{\overline{\alpha }} \, ]
  $ have the opposite directions relative to the directions of $j_0^{\alpha}$
 In this way, the corrections of the first type, increasing  the current, should be treated
 as a precursor of forming the Poiseuille viscous flow, while the second type corrections, decreasing the current and increasing the sample resistance, are a precursor of forming of a homogeneous Ohmic flow.

Using the method described in  of the previous section,
 we obtained  the correction to
the total current in zero magnetic
 field:
 \begin{equation}
 \label{delta_I_h}
\delta I_ h =  \frac{4 e_0^2  E_0   \Gamma  W^3  }{\pi^2 m }
    \, \big( n_0^h - n_0^e  \big)^2 \,
    \ln^2 \Big(\frac{1}{\gamma W}\Big)
\:,
  \end{equation}
  the correction to the Hall electric field:
 \begin{equation}
 \Delta E_H
 =   \frac{4 \omega_c \Gamma W^2 E_0  }{\pi^2} \,
 (n_0^h - n_0^e)
 \, \ln ^2 \Big(  \frac{1}{\gamma W} \Big)
\:,
  \end{equation}
  and the correction to the magnetic-field
  dependent part of the total
  current:
 \begin{equation}
 \label{Delta_I_2}
 \Delta I_2 =
\frac{6 e_0^2   E_0 \omega_c ^2 \Gamma  W }{\pi ^2 m \gamma^4 }
  \, \big( n_0^h - n_0^e  \big)^2 \,
  \ln  \Big(  \frac{1}{\gamma W} \Big)
  \:.
  \end{equation}
It is noteworthy that  all the obtained correction vanish
 at the charge neutrality point when  $ \mu_0 = 0$ and $n_0^h = n_0^e$.
and the corrections to the current
(\ref{delta_I_h}) and (\ref{Delta_I_2})
are always positive. This means that  the electron-electron and electron-hole collisions
in a non-degenerate symmetric two-component system lead together to
the hydrodynamic (but not to the Ohmic) reconstruction of the ballistic flow.
This effect however vanishes at the charge neutrality point when these
two types of scattering compensate each other.

\section{ Acknowledgements}
 We are grateful to A.~I.~Chugunov, A.~P.~Dmitriev, M.~M.~Glazov, I.~V.~Gornyi, and
V.~Yu.~Kachorovskii  for valuable discussions as well as to
  A.~P.~Alekseeva, E.~G.~Alekseeva, I.~P.~Alekseeva,
  N.~S.~Averkiev, A.~I.~Chugunov,  I.~V.~Gornyi, M.~M.~Glazov,
  and D.~S.~Svinkin for advice and support.

 This work was supported by the Russian Fund for
Basic Research (Grants No. 16-02-01166-a and 17-02-00217-a)
 and by the grant of the Basis Foundation (Grant No. 17-14-414-1).


\begin{thebibliography}{}
\bibitem{rep}
For ease of reading, we present  in this paper the model for the ballistic transport of interaction particles
formulated in Ref.~\cite{ours_bal} as well as some of the results obtained in that work.
Herewith in this paper, unlike Ref.~\cite{ours_bal}, the definitions of the current  density $j(y)$ and the total current $I$ are given with the dimensional coefficients: the unperturbed particle density $n_0$, the particl charge $e$ and the mass $m$.
\bibitem{n}
Here the expression for $E_H$ is written with an exact numeric factor,
 while the expression for magnetoresistance is written only up to an order of magnitude.
 The reason of this difference is as follows.
 The expression for $E_H$ is obtained by integrating
 of the contributions from electrons with the angles
 $\varphi $ close to $\pm \pi /2$ and
 having a logarithmic divergence by $||\varphi | - \pi/2 |$,
 while the expression for $\Delta I$ (and thus for $\Delta I_2$)
 is obtained by integrating of the contributions with the divergence
 at the same angles $\varphi \approx \pm p/2$,
  but having a stronger (not logarithmic) power singularity.
  The last does not allow  to determine (when truncating the integration
  at  $||\varphi | - \pi/2 | \sim \delta _m $)
  the numerical prefactor in  $\Delta I_2$
  and, thus, in $\Delta \Delta R $.
\end{thebibliography}
\end{document}